\documentclass[12pt,epsfig]{article}
\usepackage[X2,T2A]{fontenc}
\usepackage[koi8-r]{inputenc}
\usepackage[english]{babel}
\usepackage{graphicx}
\usepackage{amssymb}
 
\begin{document}
\begin{center}
\begin{Large}
Two-Photon 2s$\leftrightarrow$1s Transitions during Recombination of 
Hydrogen in the Universe
\end{Large}
\end{center}
\bigskip
\begin{center}
E. E. Kholupenko$^*$ and A. V. Ivanchik 
\\Ioffe Physical-Technical Institute, St. Petersburg
\end{center}
\bigskip
\begin{center} 
\begin{large} 
{\bf Abstract}
\end{large}
\end{center}
Based on the standard cosmological model, we calculate the correction to 
the rate of two-photon $2s \leftrightarrow 1s$ transitions in the hydrogen 
atom under primordial hydrogen plasma recombination conditions that arises 
when the induced transitions under equilibrium background radiation with a 
blackbody spectrum and plasma recombination radiation are taken into account. 
\vfill $^*$ eugene@astro.ioffe.ru

\newpage

\section{Introduction}
The cosmological plasma recombination plays an 
important role in producing the cosmic microwave 
background radiation (CMBR) anisotropy, since 
it determines the time when the optical depth of 
the Universe for Compton scattering becomes less 
than unity. The recombination model suggested by 
Zel'dovich et al. (1968) and Peebles (1968) for the 
hydrogen plasma and generalized by Seager et al. 
(1999, 2000) to the hydrogen-helium plasma allows 
an accuracy of at least 1\% to be achieved when calculating 
the cosmological recombination. Therefore, 
in the last few years, the efforts of researchers in 
this field have been directed at studying the effects 
that can affect the primordial plasma recombination 
kinetics at a level of up to several percent (Leung 
et al. 2004; Dubrovich and Grachev 2005; Chluba 
and Sunyaev 2006; Novosyadlyj 2006). 

The recombination kinetics depends mainly on 
two processes: the two-photon $2s \leftrightarrow 1s$ transition 
and the Ly-$\alpha$ photon escape from the $2p \rightarrow 1s$ line 
profile due to the cosmological redshift (Zel'dovich 
et al. 1968; Peebles 1968; Seager et al. 1999, 2000). 

In this paper, we discuss the correction to the rate 
of two-photon $2s \leftrightarrow 1s$ transitions due to the effect 
of induced transitions. Such an effect was first discussed 
by Chluba and Sunyaev (2006), who showed 
that including it leads to the relative change in the 
ionization fraction by one percent for epochs $z=900-1200$. 
This is important, since, in turn, it can 
lead to changes in the CMBR temperature fluctuation 
spectrum, which can be measured in planned experiments 
(Planck and others). 

In this paper, special emphasis was placed on 
studying the $1s \rightarrow 2s$ transitions ($h\nu_{21}\simeq 10.2$ eV)
related to two-photon absorption. In addition to the 
study by Chluba and Sunyaev (2006), we took into 
account the fact that the nonequilibrium Ly-$\alpha$ photons 
produced under primordial hydrogen plasma recombination 
are shifted to the lower frequencies. For the 
recombination period, the latter are limited by the frequency 
$\nu^{red}_{21}=\nu_{21}(1+z_{end})/(1+z_{begin})\simeq 0.6\nu_{21}$, 
where $z_{begin}\simeq 1600$ and $z_{end}\simeq 900$ define the period 
in which more than 95\% of the recombination 
Ly-$\alpha$ photons were produced. As a result of this 
process, the occupation number of the primordial 
plasma recombination radiation photons (reddened 
Ly-$\alpha$ photons) at the recombination epoch in the 
frequency range [$0.6\nu_{21}; \nu_{21}$] exceeds the occupation 
number of the equilibrium background photons by 1-4 orders of 
magnitude (see Fig. 1). This, in turn, results in a 
significant difference between the $1s \rightarrow 2s$ reaction 
rate and that from Chluba and Sunyaev (2006), who 
assumed the presence of only equilibrium background 
radiation with a blackbody spectrum in the above 
frequency range. 

\section{Uncompensated $2s \leftrightarrow 1s$ transitions}
Calculating the correction to the rate of two-photon 
$2s\leftrightarrow 1s$ transitions is based on calculating 
the uncompensated electron flow $J_{2s1s}$ between 
the 2s and 1s levels. The uncompensated electron 
flow $J_{2s1s}$ is described by the expression 
\begin{equation}
J_{2s1s}=P_{2s1s}x_{2s}-P_{1s2s}x_{1s}
\label{flow1}
\end{equation}
Here, $P_{2s1s}$ [s$^{-1}$] is the rate of two-photon $2s \rightarrow 1s$ 
transitions including the spontaneous and induced 
transitions, $P_{1s2s}$ is the rate of two-photon $1s \rightarrow 2s$ 
transitions including the induced transitions, $x_{2s}$ and 
$x_{1s}$ are the relative populations of the 2s and 1s states 
defined by the expressions 
\begin{equation}
x_{2s}=N_{2s}/N_{H},~~~~~x_{1s}=N_{1s}/N_{H}
\end{equation}
where $N_{2s}$ is the number density of 2s-state atoms, 
$N_{1s}$ is the number density of ground-state atoms, and 
$N_H$ is the total number density of neutral and ionized 
hydrogen atoms. 

Let us define the rate of uncompensated $2s \rightarrow 1s$ 
transitions $A^{u}_{2s1s}$ (the superscript $u$ stands for uncompensated) as 
\begin{equation}
A^{u}_{2s1s}=J_{2s1s}/x_{2s}=P_{2s1s}-P_{1s2s}{x_{1s} \over x_{2s}},
\label{A_u_definition}
\end{equation}
$A^{u}_{2s1s}$ can be comparable to the probability 
$A_{2s1s}\simeq 8.22$ s$^{-1}$ of spontaneous two-photon $2s \rightarrow 1s$ 
decays that determines the rate of $2s \rightarrow 1s$ transitions 
in the absence of an external radiation field 
(i.e., in a vacuum). In turn, the probability $A_{2s1s}$ 
of spontaneous $2s \rightarrow 1s$ transitions in the hydrogen 
atom is defined by the expression (Spitzer and Greenstein 
1951; Zon and Rapoport 1968) 
\begin{equation}
A_{2s1s}={A_0 \over 2}\int_0^{1}\phi(y)dy
\label{spont_trans1}
\end{equation}
where $A_0=9\alpha^6 c$Ry$/2^{10} \simeq 4.3663$ s$^{-1}$ is the dimensional 
normalization constant, $\alpha$ is the fine-structure 
constant, Ry is the Rydberg constant for 
hydrogen, $c$ is the speed of light, and the function 
$\phi(\nu/\nu_{21})$ describes the spontaneous two-photon decay 
spectrum (Spitzer and Greenstein 1951; Zon and 
Rapoport 1968). The value of $\phi(y)dy$ is proportional 
to the spontaneous emission probability of a pair of 
photons one of which has frequency $\nu$ in the range 
$[\nu_{21}y;\nu_{21}y+\nu_{21}dy]$ and the other has frequency 
$\nu'=(\nu_{21}-\nu)$. The factor 1/2 arises, because each pair of 
photons is taken into account twice when integrating 
in (4). 

Calculating the rate of uncompensated transitions 
$A^{u}_{2s1s}$ requires knowledge of $P_{2s1s}$ and $P_{1s2s}$. 
The transition rates $P_{2s1s}$ and $P_{1s2s}$ depend on the parameters 
of the radiation field with which the primordial 
plasma interacts. This dependence is taken into account 
by the expressions (Rapoport et al. 1978) 
\begin{equation}
P_{2s1s}={A_{0} \over 2}
\int_0^{1}\phi\left({\nu \over \nu_{21}}\right)
\left(1+\eta(\nu)\right)\left(1+\eta(\nu')\right)
d\left({\nu \over \nu_{21}}\right),
\label{2s_1s_trans}
\end{equation}
\begin{equation}
P_{1s2s}={A_{0} \over 2}
\int_0^{1}\phi\left({\nu \over \nu_{21}}\right)
\eta(\nu)\eta(\nu')d\left({\nu \over \nu_{21}}\right),
\label{1s_2s_trans}
\end{equation}
where $\eta(\nu)$ is the occupation number at frequency $\nu$ 
in the radiation field with which the primordial plasma 
interacts. 

Thus, taking Eqs. (3)-(6) into consideration, we 
can represent $A^{u}_{2s1s}$ in the following form containing 
an explicit dependence on the occupation numbers $\eta$: 
\begin{equation}
A^{u}_{2s1s}={A_{0} \over 2}
\int_0^{1}\phi\left({\nu \over \nu_{21}}\right)
\left[\left(1+\eta(\nu)\right)
\left(1+\eta(\nu')\right)-\eta(\nu)\eta(\nu')
{x_{1s} \over x_{2s}}
\right]d\left({\nu \over \nu_{21}}\right),
\label{flow2_a}
\end{equation}
Basically, Eq. (7) is the collision integral for photons 
and hydrogen atoms. 

To simplify the subsequent calculations, let us 
pass to the following expression for the rate of uncompensated 
transitions with integration over half 
the range of the $2s \rightarrow 1s$ transition, which is valid, 
since the integrands are symmetric with respect to the 
substitution $\nu \leftrightarrow (\nu_{21}-\nu)$: 
\begin{equation}
A^{u}_{2s1s}={A_{0}}
\int_0^{1/2}\phi\left({\nu \over \nu_{21}}\right)
\left[\left(1+\eta(\nu)\right)
\left(1+\eta(\nu')\right)-\eta(\nu)\eta(\nu')
{x_{1s} \over x_{2s}}
\right]d\left({\nu \over \nu_{21}}\right),
\label{flow2}
\end{equation}

Although the integration in Eq. (8) is over the frequency 
range $[0;\nu_{21}/2]$, the argument $\nu'=(\nu_{21}-\nu)$ 
varies over the range $[\nu_{21}/2;\nu_{21}]$, i.e., the integrand 
depends on the occupation numbers $\eta$ in the full 
transition frequency range $[0;\nu_{21}]$. 

Since the frequency of more than 95\% of the Ly-$\alpha$ 
photons does not decrease to $\nu_{21}/2$ in the recombination 
period and since the perturbations of the 
occupation numbers related to the transitions between 
excited states of the hydrogen atom (transition 
frequencies $<\nu_{21}/2$) are no larger than $10^{-5}$ (see, 
e.g., Dubrovich and Grachev 2004; Kholupenko et al. 
2005), the occupation number $\eta(\nu)$ in the frequency 
range $0\le \nu \le \nu_{21}/2$ is described with a relative accuracy 
of at least $10^{-3}$ by the Planck distribution 
\begin{equation}
\eta^0(\nu)=\left( \exp\left({h\nu \over k_BT}\right)-1 \right)^{-1}
\label{Planck1}
\end{equation} 
where $k_B$ is the Boltzmann constant, the parameter 
$T=T_0(1+z)$ is the CMBR temperature at epoch $z$, 
and $T_0=2.726$ K is the CMBR temperature at the 
present epoch. Accordingly, the following approximate 
equality is valid for $A^{u}_{2s1s}$: 
\begin{equation}
A^{u}_{2s1s}={A_{0}}
\int_0^{1/2}\phi\left({\nu \over \nu_{21}}\right)
\left[\left(1+\eta^0(\nu)\right)
\left(1+\eta(\nu')\right)-\eta^0(\nu)\eta(\nu')
{x_{1s} \over x_{2s}}
\right]d\left({\nu \over \nu_{21}}\right).
\label{A_unc_2}
\end{equation}
The second approximation that allows Eq. (8) to be 
simplified stems from the fact that the occupation 
numbers $\eta (\nu')$ are no larger in absolute value than 
$10^{-5}$ (i.e., $\eta (\nu') \ll 1$) in the frequency range 
$\nu_{21}/2\le \nu'\le \nu_{21}$
at the recombination epoch. Therefore, $\eta (\nu')$
in the first term in the integrand in (10) may be 
ignored compared to unity. At the same time, $\eta (\nu')$
cannot be ignored in the second term, since it appears 
in Eq. (10) in the product with a large factor 
$x_{1s}/x_{2s} \gg 1$. Taking this into account, we obtain a 
simplified expression for $A^{u}_{2s1s}$: 
\begin{equation}
A^{u}_{2s1s}={A_{0}}
\int_0^{1/2}\phi\left({\nu \over \nu_{21}}\right)
\left[\left(1+\eta^0(\nu)\right)-\eta^0(\nu)\eta(\nu')
{x_{1s} \over x_{2s}}
\right]d\left({\nu \over \nu_{21}}\right).
\label{A_unc_3}
\end{equation} 

Let us introduce the function 
\begin{equation}
R(\nu) \equiv \left(1-{\eta(\nu')\eta^0(\nu) 
\over 1+\eta^0(\nu)}
{x_{1s} \over x_{2s}}\right),
\label{R_definition}
\end{equation}
that characterizes the $2s \leftrightarrow 1s$ reaction rate at certain 
emitted photon frequencies $\nu$ and $\nu'=(\nu_{21}-\nu)$. If 
the rate of the reverse reaction is negligible compared 
to the rate of the direct reaction (e.g., for late epochs 
$z \lesssim 1000$), then $R \simeq 1$. If, alternatively, the direct 
and reverse reactions compensate each other (e.g., at 
early epochs $z \gtrsim 1500$ when the populations are close 
to the equilibrium ones), then $R \simeq 0$. 

Using definition (12), we can write Eq. (11) in a 
compact form:
\begin{equation}
A^{u}_{2s1s}={A_{0}}\int_0^{1/2}\phi
\left({\nu \over \nu_{21}}\right)\left(1+\eta^0(\nu)\right)
R (\nu) d\left({\nu \over \nu_{21}}\right)
\label{A_compact}
\end{equation}

The optical depth in the Ly-$\alpha$ line at the recombination 
epoch is $10^5-10^8$, which leads to the following 
relation between the populations $x_{1s}, x_{2p}$ and the occupation 
number in the Ly$\alpha$ line: $\eta(\nu_{21})={x_{2p} / 3x_{1s}}$. 
Since the condition $x_{2p}=3x_{2s}$ is satisfied with a 
relative accuracy of at least $10^{-2}$ for epochs $z \ge 800$ 
(Grachev and Dubrovich 1991), we obtain the following 
expression for $\eta(\nu_{21})$ (see also Peebles 1968): 
\begin{equation}
\eta(\nu_{21})={x_{2s} \over x_{1s}}
\label{Peebles_determine1}
\end{equation}
Substituting Eqs. (9) and (14) in Eq. (12) yields 
\begin{equation}
R(\nu) = \left(1-{\eta(\nu') \over \eta(\nu_{21})}
\exp\left(-{h\nu \over k_BT}\right)\right)
\label{R2}
\end{equation}

The occupation number $\eta(\nu_{21})$ in the Ly-$\alpha$ line 
can be calculated using the recombination equation 
(Peebles 1968). In our notation, this equation is 
\begin{equation}
\dot x_p = - \alpha_c N_H x_p^2 + \beta_c \eta (\nu_{21}) (1-x_p)
\label{rec_eq1}
\end{equation}
where $x_p=N_p/N_H$ is the hydrogen plasma ionization 
fraction, $\alpha_c$ is the total recombination coefficient 
to all excited states of the hydrogen atom, and $\beta_c$ is 
the total ionization coefficient from all excited states. 
These coefficients are related by 
\begin{equation}
\beta_c=\alpha_c g_e(T) \exp\left({-{E_2 \over k_B T}}\right) 
\label{detailed_balance}
\end{equation}
where $g_e(T)$ is the partition function of free electrons 
and $E_2$ is the ionization energy of the hydrogen atom 
in the 2s state. 

Using (16) and (17), we can write 
\begin{equation}
\eta(\nu_{21})=\gamma(z)\eta^0(\nu_{21})(1+\delta (z)),~~~~
\delta (z)={\dot x_p \over \alpha_c N_H x_p^2}
\end{equation}
where $\gamma (z)$ is the function that characterizes the 
plasma deviation from ionization equilibrium, 
\begin{equation}
\gamma (z)={x_p^2 \over (1-x_p)}{N_{H} \over g_e(T)}
\exp\left({E_1 \over k_BT}\right)
\label{overheating}
\end{equation}
where $E_1$ is the ionization energy of the hydrogen 
atom in the ground state. Calculating the function 
$\gamma (z)$ requires prespecifying the dependence $x_p(z)$, 
which is determined by solving the problem of primordial 
plasma recombination through numerical calculation 
(e.g., using {\bf recfast}; Seager et al. 1999). The 
calculation of the recombination kinetics itself may 
be performed without including the effect considered 
here, i.e., in terms of the standard recombination 
model (Peebles 1968; Seager et al. 1999, 2000), since 
the expected relative change in the ionization fraction 
after applying the correction under study to the 
rate of two-photon $2s \leftrightarrow 1s$ decay is $\sim 10^{-2}$ 
(Chluba and Sunyaev 2006). The function $\gamma (z)$ is presented 
in Fig. 2. In this paper, we will consider $\gamma (z)$ as a 
known quantity in terms of which all of the soughtfor 
functions will be expressed. 

Since the value of the function $|\delta (z)|$ for epochs 
$z \ge 1100$ does not exceed $2 \cdot 10^{-2}$, below we will use 
an approximate equality to determine $\eta(\nu_{21})$ (disregarding 
$\delta (z)$ compared to unity): 
\begin{equation}
\eta(\nu_{21},z)\simeq \gamma(z)\eta^0(\nu_{21},z)
\label{Peebles_determine2}
\end{equation}

The last function to be determined to calculate 
$R(\nu)$ is the total occupation number $\eta (\nu')$ of CMBR 
photons and recombination Ly-$\alpha$ photons whose frequency 
decreased from the initial value of $\nu_{21}$ to $\nu'$ as 
\begin{equation}
\nu'=\nu_{21}{1+z \over 1+z'}
\label{freq_change}
\end{equation}
where $z'$ is the epoch when the photon frequency 
was $\nu_{21}$. Taking into account the law of frequency 
change (21), we obtain $\eta(\nu', z)=\eta(\nu_{21},z')$ and 
$\eta^0(\nu', z)=\eta^0(\nu_{21},z')$. Combining these equalities 
and equality (20) leads to the following formulae to 
determine $\eta (\nu', z)$: 
\begin{equation}
\eta (\nu', z)=\eta (\nu_{21},z') 
=\eta^0(\nu_{21},z') \gamma (z') = \eta^0(\nu', z) \gamma (z') 
\label{Ly_expression}
\end{equation}
where the parameter $z'$ depends on $\nu'$ and $z$ and is 
defined by the inequality that follows from (21): 
\begin{equation}
z'(\nu',z)=\left({\nu_{21} \over \nu'}\right)(1+z)-1
\end{equation}

Given (20) and (22), the formula for $R(\nu)$ can be 
written as 
\begin{equation}
R (\nu)=\left(1-{\gamma (z') \over \gamma (z)}
{\eta^0(\nu') \over \eta^0(\nu_{21})}
\exp\left(-{h\nu \over k_BT}\right)
\right)
\label{R3}
\end{equation}

A further simplification of this formula is related 
to the fact that the following equality is valid for 
$\nu' \ge \nu_{21}/2$ at the recombination epoch: 
\begin{equation}
{\eta^0(\nu') \over \eta^0(\nu_{21})}
\exp\left(-{h\nu \over k_BT}\right)\simeq 1
\end{equation}
since the approximate equality 
\begin{equation}
\eta^0(\nu') \simeq \exp\left(-{h\nu' \over k_BT}\right),  
\end{equation}
holds with a relative accuracy of at least $10^{-5}$ in this 
period. Taking this into account, we obtain the final 
formula for $R(\nu)$: 
\begin{equation}
R (\nu)=\left(1-{\gamma (z') / \gamma (z)}\right)
\label{R4}
\end{equation}
Substituting this expression in Eq. (13) yields the 
final expression for the rate of uncompensated decays 
\begin{equation}
A^{u}_{2s1s}={A_{0}}\int_0^{1/2}\phi
\left({\nu \over \nu_{21}}\right)\left(1+\eta^0(\nu)\right)
\left(1-{\gamma (z') / \gamma (z)}\right)d\left({\nu \over \nu_{21}}\right)
\end{equation}
If equilibrium radiation is assumed to be present in the 
frequency range $[0;\nu_{21}]$, then the principle of detailed 
balancing can be used (Chluba and Sunyaev 2006). 
The rate of uncompensated transitions calculated under 
this assumption is given by the expression (in 
addition, highlighted by the superscript $CS$): 
\begin{equation}
A^{u,CS}_{2s1s}={A_{0}}\int_0^{1/2}\phi
\left({\nu \over \nu_{21}}\right)\left(1+\eta^0(\nu,z)\right)
\left(1-1 / \gamma (z)\right) d\left({\nu \over \nu_{21}}\right)
\label{A_ueq}
\end{equation}

In the standard recombination calculation (Peebles 
1968; Seager et al. 1999, 2000), no induced transitions 
are considered, i.e., the occupation number 
$\eta^0(\nu)$ in Eq. (29) is disregarded. Thus, the rate of 
uncompensated transitions can be calculated using 
the formula (the superscript $st$ stands for standard) 
\begin{equation}
A^{u,st}_{2s1s}={A_{0} }\int_0^{1/2}\phi
\left({\nu \over \nu_{21}}\right)\left(1-1 / \gamma (z)\right) 
d \left({\nu \over \nu_{21}}\right).
\label{A_ust}
\end{equation}

Since the key aspect of this study is allowance for 
the recombination Ly-$\alpha$ photons (because the $1s \rightarrow 2s$ 
reaction rate depends on their number), we calculated 
the ratio of the intensity of the ``reddened'' recombination 
radiation to the intensity of the equilibrium background radiation with 
temperature $T_0$ that must be observed at the current 
epoch (Fig. 3). This quantity is related to the function $\gamma(z)$ by
\begin{equation}
{I(z=0)_{distortion} \over I(z=0)_{Planck}}=
{\eta(\nu)-\eta^0(\nu) \over \eta^0(\nu)} = 
\gamma(z_\nu)-1
\end{equation}
where $z_{\nu}=\nu_{21}/\nu-1$ is the redshift of the Ly-$\alpha$ photon 
produced at the recombination epoch. Previously, 
this quantity was calculated by Wong et al. (2005). 
Our calculations show satisfactory agreement with 
the calculations by Wong et al. (2005). Our calculations 
also agree with those of Grachev and 
Dubrovich (1991) and Boschan and Biltzinger (1998), 
who calculated the distortion of the CMBR intensity 
related to the recombination radiation in the Ly-$\alpha$ line. 

\section{Results}
The calculations were performed in terms of the 
standard $\Lambda$CDM cosmological model with the following 
parameters at the present epoch: the Hubble 
constant $H_0=70$ km/s/Mpc; the relative 
vacuum-like energy density (cosmological constant) 
$\Omega _\Lambda \simeq 0.73$; the relative nonrelativistic matter 
density $\Omega _m \simeq 0.27$; the relative baryonic density 
$\Omega_b \simeq 0.036$; and the relative relativistic matter density 
$\Omega_{rel} \simeq 10^{-4}$ (photons + neutrinos); 

Figure 4 reproduces the results by Chluba and 
Sunyaev (2006): the spectrum (solid curve) of the 
spontaneous two-photon $2s \rightarrow 1s$ decay $\phi (\nu / \nu_{21})$; 
and the total spectrum (dotted curve) of the spontaneous 
and induced two-photon $2s \rightarrow 1s$ decays 
(Chluba and Sunyaev 2006), which is symmetric 
with respect to $\nu_{21}/2$ and is defined by the following 
function in the range $[0;\nu_{21}/2]$: 
\begin{equation}
\phi^{ind}\left({\nu \over \nu_{21}}\right)=
\phi \left({\nu \over \nu_{21}}\right)
\left(1+\eta^0(\nu)\right), 
\end{equation}
calculated for the epoch $z=1500$ (the superscript $ind$ 
stands for induced). 

Figure 5 presents the following calculated spectra: 

(1) The spectrum of the two-photon $2s \leftrightarrow 1s$ transition 
including only the spontaneous decays, with 
the reverse reaction channel being taken into account 
using the principle of detailed balancing: 
\begin{equation}
\phi^{st}\left({\nu \over \nu_{21}}\right)=\phi
\left({\nu \over \nu_{21}}\right) \left(1-1 / \gamma (z)\right) 
\end{equation}
This spectrum was calculated under the assumptions 
made in the standard recombination model (Peebles 
1968; Seager et al. 1999). In Fig. 5, it is represented 
by the solid curve. 

(2) The spectrum of the two-photon $2s \leftrightarrow 1s$ transition 
including the spontaneous and induced transitions, 
the direct and reverse reaction channels, and 
under the assumption that the radiation is equilibrium 
one in the entire $2s \rightarrow 1s$ transition frequency 
range: $\phi^{CS}\left({\nu / \nu_{21}}\right)$. 
This spectrum is also symmetric 
with respect to $\nu_{21}/2$ and is defined by the following 
function in the range $[0;\nu_{21}/2]$: 
\begin{equation}
\phi^{CS}\left({\nu \over \nu_{21}}\right)=\phi
\left({\nu \over \nu_{21}}\right) 
\left(1+\eta^0(\nu)\right) \left(1-1 / \gamma (z)\right) 
\end{equation}
This spectrum was calculated under the assumptions 
made by Chluba and Sunyaev (2006). In Fig. 5, it is 
represented by the dotted curve. 
(3) The spectrum of the two-photon $2s \leftrightarrow 1s$ transition 
including the spontaneous and induced transitions, 
the direct and reverse reaction channels, and 
the nonequilibrium radiation in the Ly-$\alpha$ line and the 
range $0.6\nu_{21}\le \nu' \le \nu_{21}$. This spectrum is 
also symmetric with respect to $\nu_{21}/2$ and is defined by 
the following function in the range $[0;\nu_{21}/2]$: 
\begin{equation}
\phi^u \left({\nu \over \nu_{21}} \right) = 
\phi \left({\nu \over \nu_{21}}\right)
\left(1+\eta^0(\nu)\right) R (\nu)
\label{phi_unc_definition}
\end{equation}
This spectrum was calculated under the assumptions 
made here. In Fig. 5, it is represented by the dashed 
curve. 

In contrast to the frequency-integrated reaction 
rates $A^{u,st}_{2s1s},~A^{u,CS}_{2s1s}$, and $A^{u}_{2s1s}$ 
the functions $\phi^{st},~\phi^{CS}$, and $\phi^u$ 
presented in Fig. 5 characterize the behavior of 
the $2s\leftrightarrow 1s$ reaction rate not only as a function of the 
epoch ($z=1500$ - Fig. 5, top; $z=1100$ - Fig. 5, bottom), but 
also as a function of the emitted photon frequencies 
$\nu$ and $\nu'=(\nu_{21}-\nu)$. The latter is important, since the 
key aspect of this paper is allowance for the effect of 
reddened recombination Ly-$\alpha$ photons, whose number 
depends significantly on frequency, on the $2s\leftrightarrow 1s$ 
reaction rate. 

Note also that the spectra $\phi^{st},~\phi^{CS}$, and $\phi^u$ shown 
in Fig. 5 are of interest in their own right when calculating 
the distortion of the CMBR by the recombination 
photons produced during two-photon $2s\leftrightarrow 1s$ transitions 
(Zel'dovich et al. 1968; Wong et al. 
2005), since these functions describe the behavior of 
the source of recombination photons for each specific 
recombination model: $\phi^{st}$ for the standard recombination 
model (Peebles 1968; Seager et al. 1999); $\phi^{CS}$ 
for the modification of the standard model suggested 
by Chluba and Sunyaev (2006); and $\phi^{u}$ for the modification 
of the standard model considered here. 

The fact that the spectrum $\phi^{u}$, in contrast to the 
spectrum $\phi^{CS}$, becomes zero at the boundary of the 
transition frequency range (at $\nu=0$) is related to 
condition (14). The latter leads to full compensation 
of the induced two-photon $2s \rightarrow 1s$ transitions by the 
induced reverse two-photon $1s \rightarrow 2s$ transitions at 
the frequency $\nu'=\nu_{21}$ that corresponds to the Ly-$\alpha$ 
photon absorption. 

Note also that the behavior of the spectra as a 
function of the epoch $z$ under consideration is peculiar 
in that for epochs $z \gtrsim 1375$ (the model-dependent 
value obtained by numerical calculation), the values 
of $\phi^{u}(\nu)$ are higher than those of $\phi^{st}(\nu)$ (see Fig. 5, top), 
i.e., $A^{u}_{2s1s}>A^{u,st}_{2s1s}$, while for epochs $z \lesssim 1375$, 
the values of $\phi^{u}(\nu)$ are lower than those of $\phi^{st}(\nu)$ (see 
Fig. 5, bottom), i.e., $A^{u}_{2s1s}<A^{u,st}_{2s1s}$. This is because the 2s 
level for $z \lesssim 1375$ is underpopulated with respect to 
the 1s level, given the occupation number $\eta(\nu')$, i.e., 
\begin{equation}
x_{2s}/x_{1s}=\eta(\nu_{21})<\eta(\nu')
\end{equation}
This, in turn, is attributable to the nonmonotonic 
behavior of $\eta(\nu')$ with frequency and epoch $z$ under 
consideration (see Fig. 6). 

Figure 7 shows the relative changes $\delta A^{CS}$ and 
$\delta A$ in the rate of uncompensated $2s\leftrightarrow 1s$ transitions 
compared to the standard calculation (Peebles 1968; 
Seager et al. 1999) related to the inclusion of induced 
transitions. The value of $\delta A^{CS}$ was calculated 
under the assumptions made by Chluba and Sunyaev 
(2006); the value of $\delta A$ was calculated under 
the assumptions made here. By definition (3), $\delta A^{CS}$ and 
$\delta A$ also describe the relative changes in the uncompensated 
electron flow $J_{2s1s}$ between the 2s 
and 1s levels compared to the standard calculation 
(Peebles 1968; Seager et al. 1999): 
\begin{equation} 
\delta A^{CS}= 
{A^{u,CS}_{2s1s} - A^{u,st}_{2s1s} \over A^{u,st}_{2s1s}}=
{J^{CS}_{2s1s} - J^{st}_{2s1s} \over J^{st}_{2s1s}}
\label{correction_CS}
\end{equation}
(dotted curve) and
\begin{equation} 
\delta A= 
{A^{u}_{2s1s} - A^{u,st}_{2s1s} \over A^{u,st}_{2s1s}}=
{J_{2s1s} - J^{st}_{2s1s} \over J^{st}_{2s1s}}
\label{correction}
\end{equation}
(dashed curve). 

Figure 8 presents the results of our calculations of 
the following quantities: 

(1) Figure 8 (top) shows the calculated total flow 
$J_{r}$ [cm$^{-3}$s$^{-1}$] of recombining primordial hydrogen 
plasma electrons: 
\begin{equation}
J_{r}=H(z)(1+z)N_H(z){dx_p \over dz},
\end{equation}
where $H(z)$ is the Hubble constant at epoch $z$. 

(2) Figure 8 (bottom) shows the relative change $\delta J_{r}$ in 
the flow of recombining electrons compared to the 
standard recombination model (Peebles 1968; Seager 
et al. 1999) related to the inclusion of induced transitions: 
\begin{equation}
\delta J_{r}={J_{r}-J^{st}_{r} \over J^{st}_{r}}.
\end{equation}

Figure 9 shows the calculated primordial hydrogen 
plasma ionization fraction $x_p$ as a function of redshift 
(Fig. 9, top) and the relative change in this quantity 
$\delta x_p=(x_p-x^{st}_p)/x^{st}_p$ compared to the standard 
recombination model (Peebles 1968; Seager et al. 
1999) related to the inclusion of induced transitions 
(Fig. 9, bottom). 

\section{Conclusions}

The calculations show that when the radiation 
produced during the primordial hydrogen plasma recombination 
and the cosmological redshift is taken 
into account, the correction $\delta A$ to the rate of uncompensated 
$2s \leftrightarrow 1s$ transitions due to the induced two-photon 
$2s \leftrightarrow 1s$ transitions is a factor of 2 smaller 
than that calculated under the assumptions made by 
Chluba and Sunyaev (2006) (see Fig. 7) and even has 
the negative sign at epochs $z \lesssim 1375$, which generally 
leads to a delay of the primordial plasma recombination 
(see Fig. 9) compared with the standard calculation 
(Peebles 1968; Seager et al. 1999). The relative 
change $\delta x_p$ in the ionization fraction for epochs $z \lesssim 1200$ 
is $0.5 \%$. This, in turn, leads to the surface of 
the last scattering being located at a lower $z$ than that 
obtained in the standard calculation of the primordial 
plasma recombination kinetics (Peebles 1968; Seager 
et al. 1999). Further calculations are required to 
accurately calculate the location of the surface of the 
last scattering and the CMBR temperature fluctuation 
spectrum by taking into account the modification 
of the recombination model suggested here. 

{\bf Acknowledgments}

We wish to thank Prof. D.A. Varshalovich for 
a detailed discussion of the paper. This work was 
supported by the Russian Foundation for Basic 
Research (project nos. 05-02-17065-a), the ``Leading 
Scientific Schools of Russia'' Program (NSh- 
9879.2006.2), and the Foundation for Support of 
Russian Science. 

\section{References} 

1. P. Boschan and P. Biltzinger, Astron. Astrophys. 336, 
1 (1998). 
\\2. J. Chluba and R. A. Sunyaev, Astron. Astrophys. 446, 
39 (2006). 
\\3. V. K. Dubrovich and S. I. Grachev, Pis'ma Astron. Zh. 
30, 723 (2004) [Astron. Lett. 30, 657 (2004)]. 
\\4. V. K. Dubrovich and S. I. Grachev, Astron. Lett. 31, 
359 (2005). 
\\5. S. I. Grachev and V. K. Dubrovich, Astrofizika 34, 249 
(1991) [Astrophys. 34, 124 (1991)]. 
\\6. E. E. Kholupenko, A. V. Ivanchik, and D. A. Varshalovich, 
Gravit. Cosmol. 11, 161 (2005). 
\\7. P. K. Leung, C. W. Chan, and M.-C. Chu, Mon. Not. 
R. Astron. Soc. 349, 632 (2004). 
\\8. B. Novosyadlyj, astro-ph/0603674 (2006). 
\\9. P. J. Peebles, Astrophys. J. 153, 1 (1968). 
\\10. L. P. Rapoport, B. A. Zon, and N. L. Manakov, Theory 
of Multiphoton Processes in Atoms (Atomizdat, Moscow, 1978) [in Russian]. 
\\11. S. Seager, D. D. Sasselov, and D. Scott, Astrophys. 
J. 523, L1 (1999). 
\\12. S. Seager, D. D. Sasselov, and D. Scott, Astrophys. 
J., Suppl. Ser. 128, 407 (2000). 
\\13. L. J. Spitzer and J. L. Greenstein, Astrophys. J. 114, 
407 (1951). 
\\14. W. Y. Wong, S. Seager, and D. Scott, astroph/0510634 (2005). 
\\15. Ya. B. Zel'dovich et al., Zh. . Eksp. Teor. Fiz. 55, 278 
(1968) [Sov. Phys. JETP 28, 146 (1968)]. 
\\16. B. A. Zon and L. P. Rapoport, Pis'ma Zh. . Eksp. Teor. 
Fiz. 7, 70 (1968) [JETP Lett. 7, 52 (1968)].

\newpage
\begin{figure}
\centering
\includegraphics[height=20cm, width=16cm]{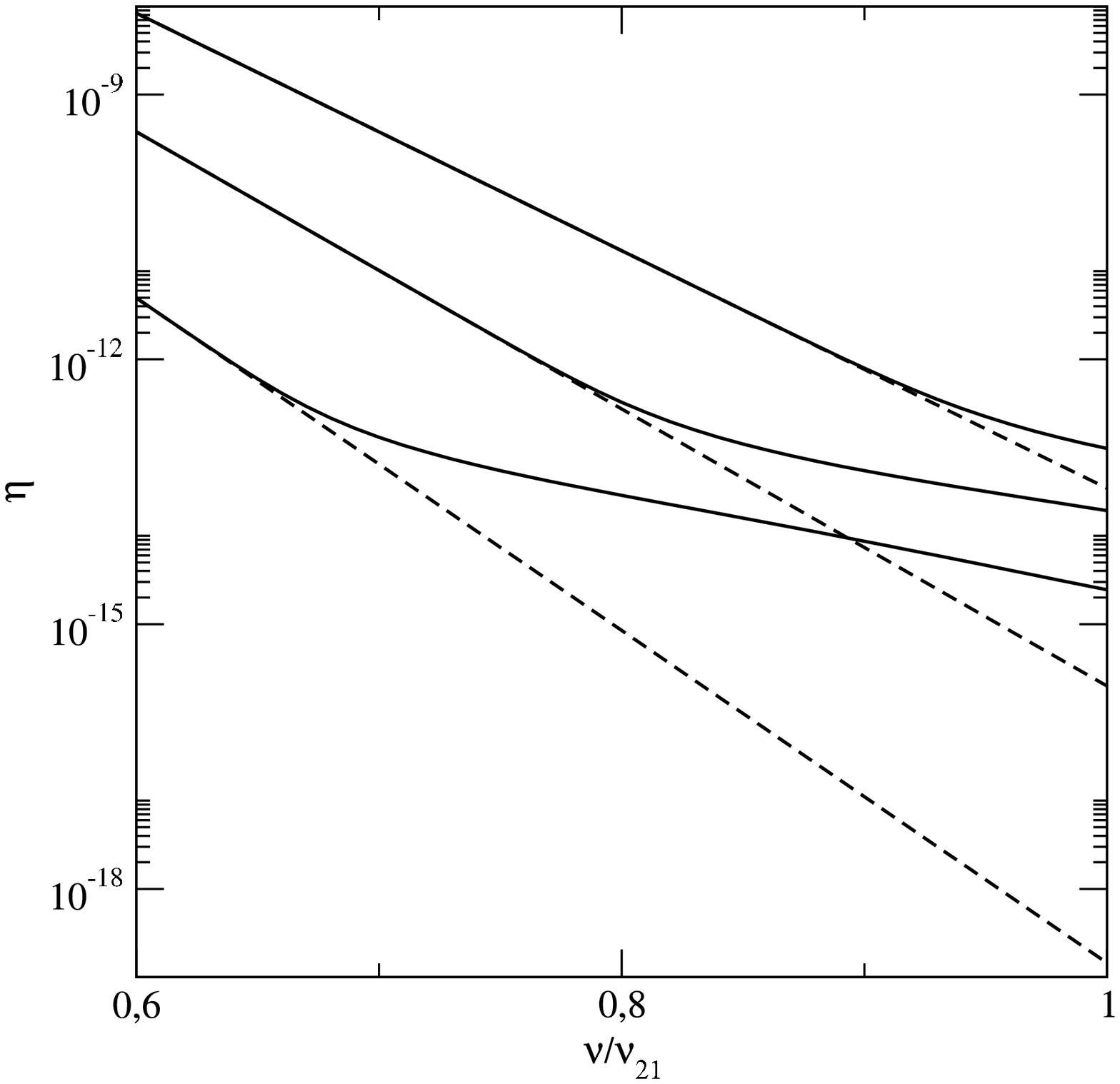}
\caption{Fig. 1. Total occupation number $\eta$ for the primordial 
hydrogen plasma recombination radiation and the 
background equilibrium radiation vs. frequency $\nu$ (solid 
curves); occupation number $\eta^0$ for the background equilibrium 
radiation vs. frequency $\nu$ (dashed curves). The excess 
of the occupation numbers $\eta$ above their equilibrium 
values of $\eta^0$ (i.e., the difference $(\eta - \eta^0)$) is determined by 
the number of ``reddened'' recombination Ly-$\alpha$ photons. 
The curves from the top downward correspond to the 
epochs $z=1400,~1200,~1000$.}
\label{graph10}
\end{figure}
 
\begin{figure}
\centering
\includegraphics[height=20cm, width=16cm]{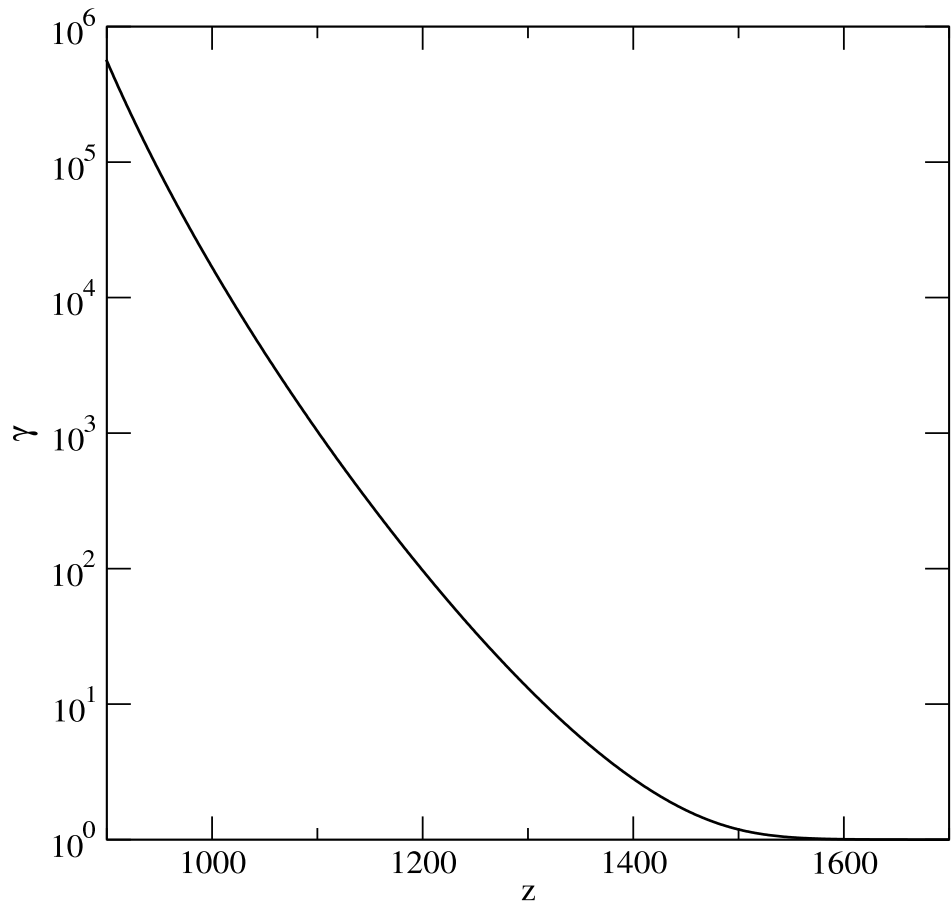}
\caption{Fig. 2. Function $\gamma (z)$ characterizing the plasma deviation 
from ionization equilibrium calculated in terms of the 
standard $\Lambda$CDM cosmological model.}
\label{graph1}
\end{figure}

\begin{figure}
\centering
\includegraphics[height=20cm, width=16cm]{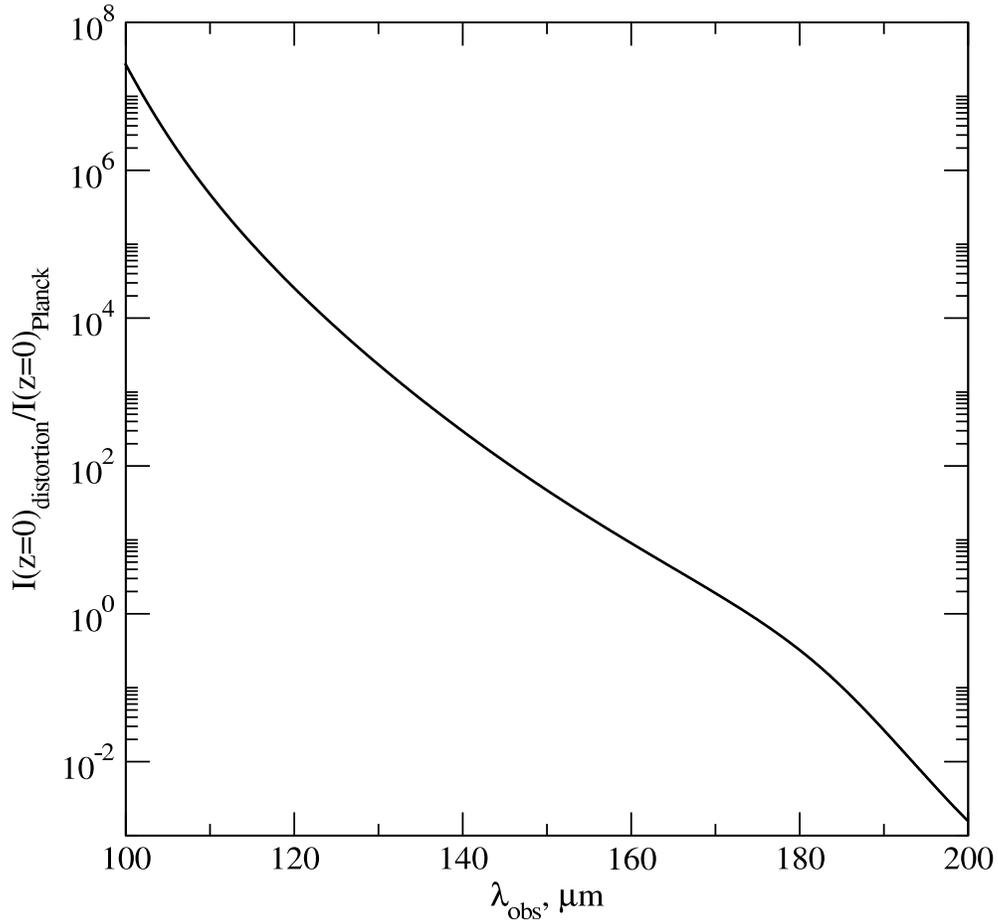}
\caption{Fig. 3. Ratio of the intensity of the primordial hydrogen 
plasma recombination radiation to the intensity of the 
CMBR ($T_0=2.726$ K) at the current epoch vs. observed 
wavelength $\lambda_{obs}$.}
\label{graph9}
\end{figure}

\begin{figure}
\centering
\includegraphics[height=20cm, width=16cm]{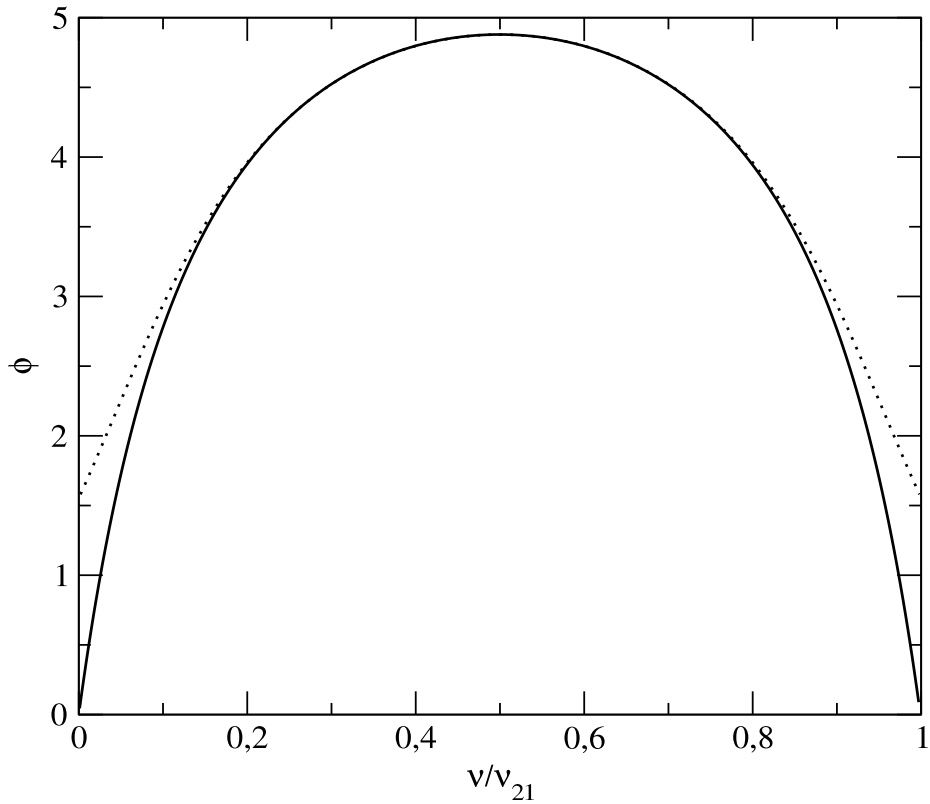}
\caption{Fig. 4. Spectrum of the spontaneous two-photon decays 
$\phi$ (solid curve) and spectrum of the two-photon 
decays including the spontaneous and induced decays 
$\phi^{ind}$ (dotted curve) for the epoch z=1500 (Chluba and 
Sunyaev 2006).}
\label{graph3}
\end{figure}

\begin{figure}
\centering
\includegraphics[width=11cm, clip, bb=32 46 575 485]{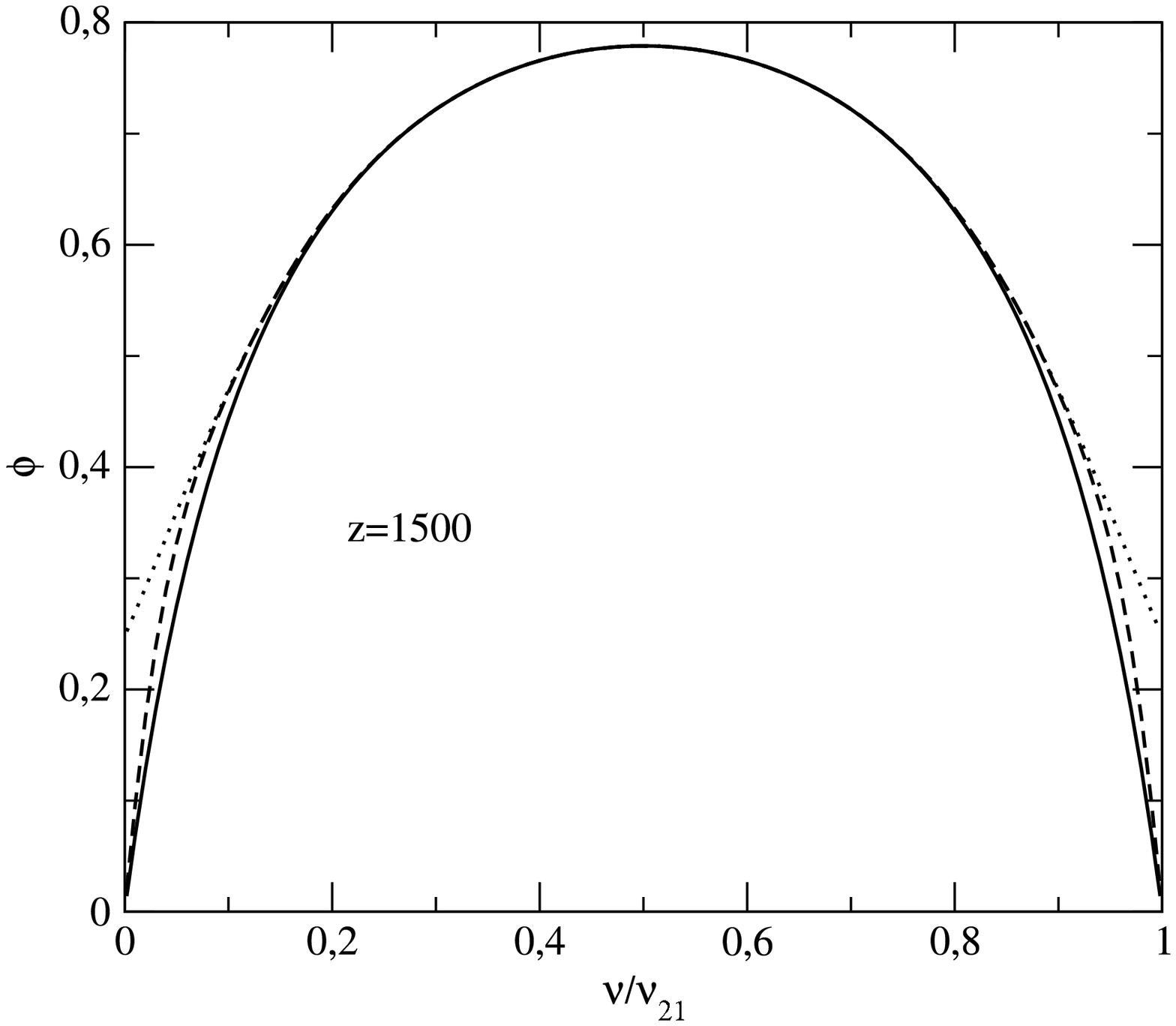}
\includegraphics[width=11cm, clip, bb=32 46 575 485]{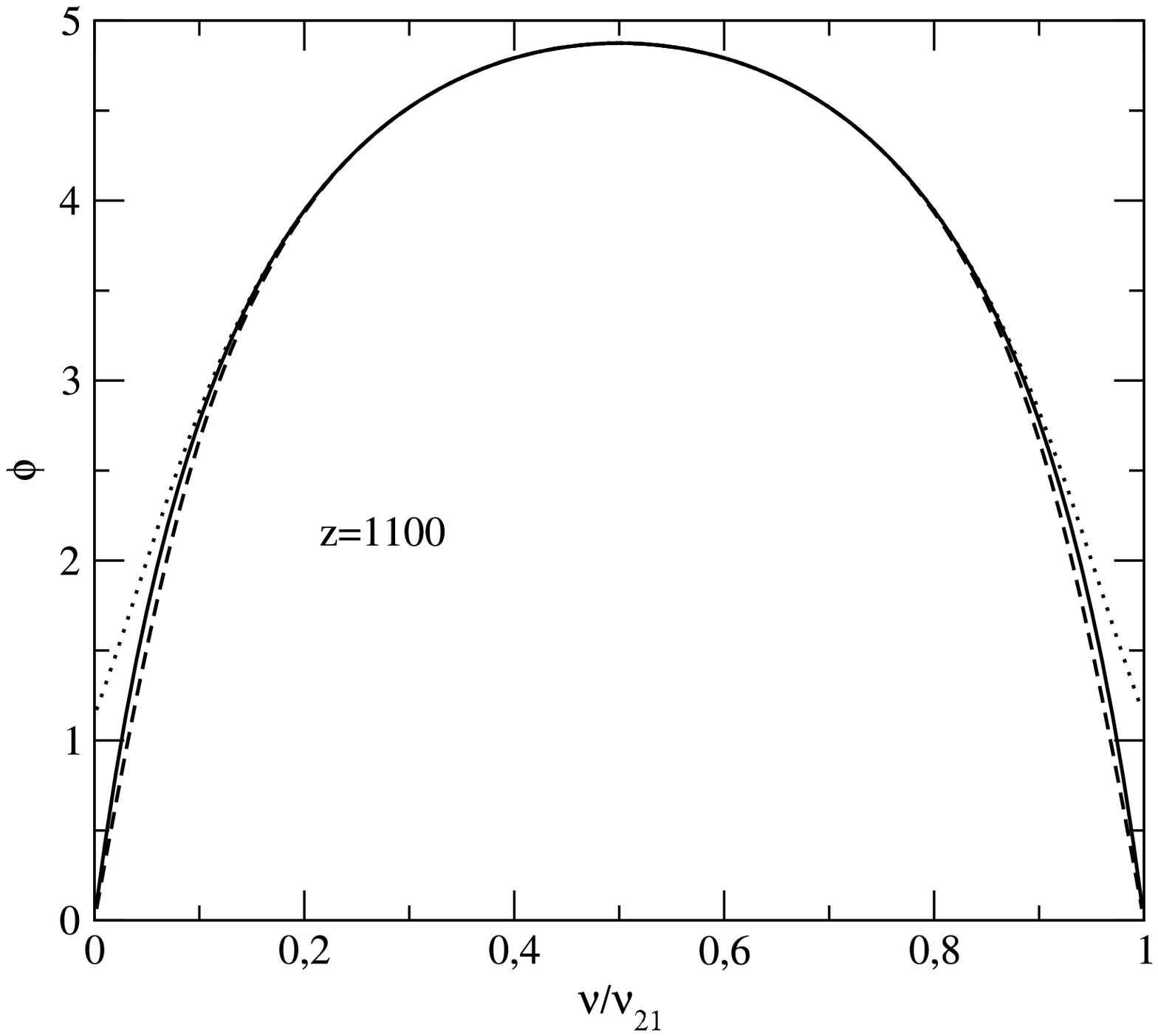}
\caption{Fig. 5. Spectra of the two-photon $2s \leftrightarrow 1s$ transitions 
at various epochs: the spectrum $\phi^{st}$ calculated under the 
assumptions of the standard recombination model (Peebles 
1968; Seager et al. 1999) (solid curve); the spectrum 
$\phi^{CS}$ calculated under the assumptions made by Chluba 
and Sunyaev (2006) (dotted curve); and the spectrum 
$\phi^{u}$ calculated under the assumptions made here (dashed 
curve).}
\label{graph6}
\end{figure}

\begin{figure}
\centering
\includegraphics[height=20cm, width=16cm]{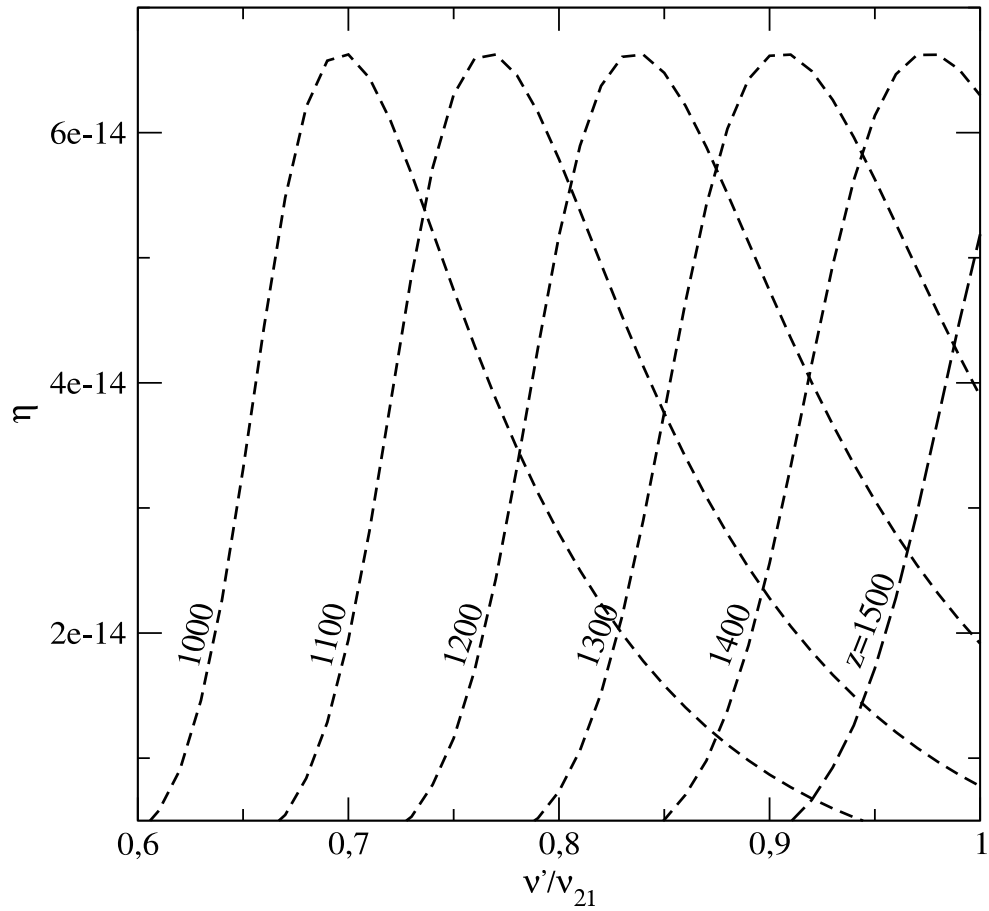}
\caption{Fig. 6. Dynamics of the recombination photon occupation 
number $\eta$: the curves for epochs $z=1000 - 1500$ are 
presented.}
\label{graph8}
\end{figure}

\begin{figure}
\centering
\includegraphics[height=20cm, width=16cm]{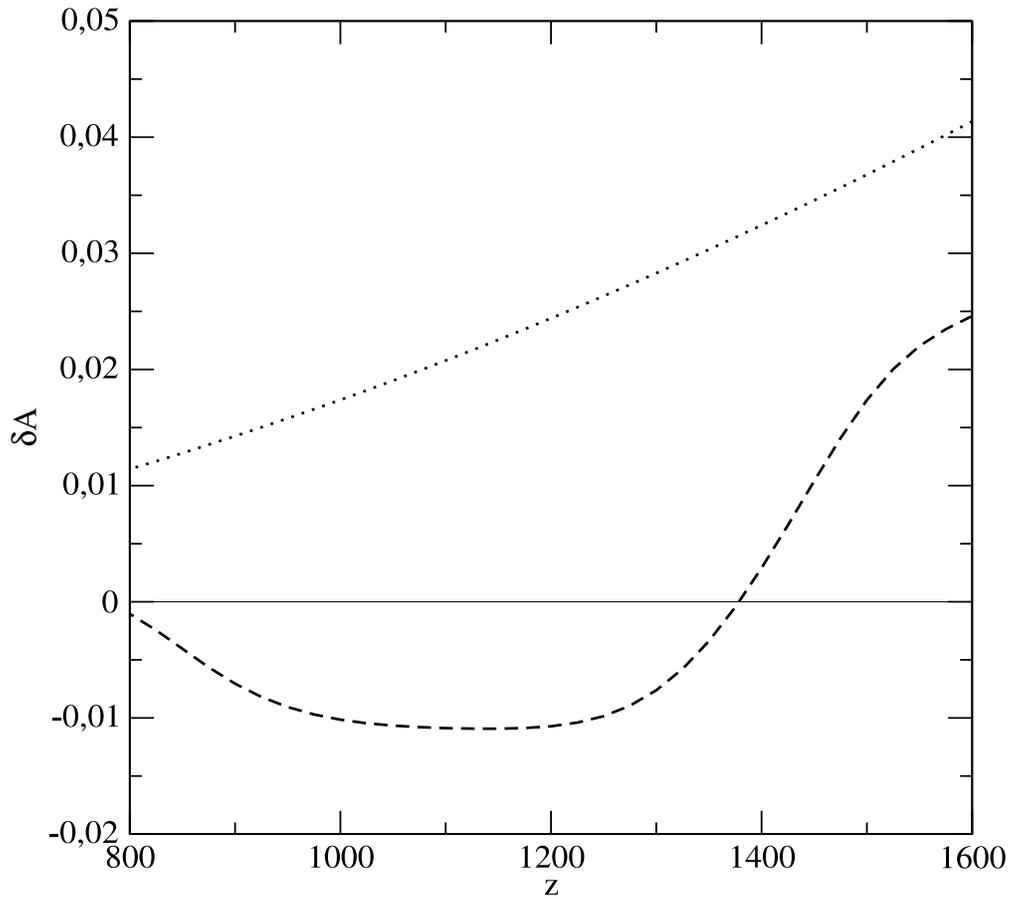}
\caption{Fig. 7. Relative change in the rate of uncompensated 
$2s\leftrightarrow 1s$ transitions compared to the rate of 
uncompensated $2s\leftrightarrow 1s$ transitions calculated under 
the assumptions 
made in the standard model (Peebles 1968; Seager et al. 1999): 
$\delta A^{CS}$  is the result obtained under the assumptions 
made by Chluba and Sunyaev (2006) (dotted curve); and $\delta A$ 
is the result obtained under the 
assumptions made here (dashed curve).}
\label{graph7}
\end{figure}

\begin{figure}
\centering
\includegraphics[height=20cm, width=16cm]{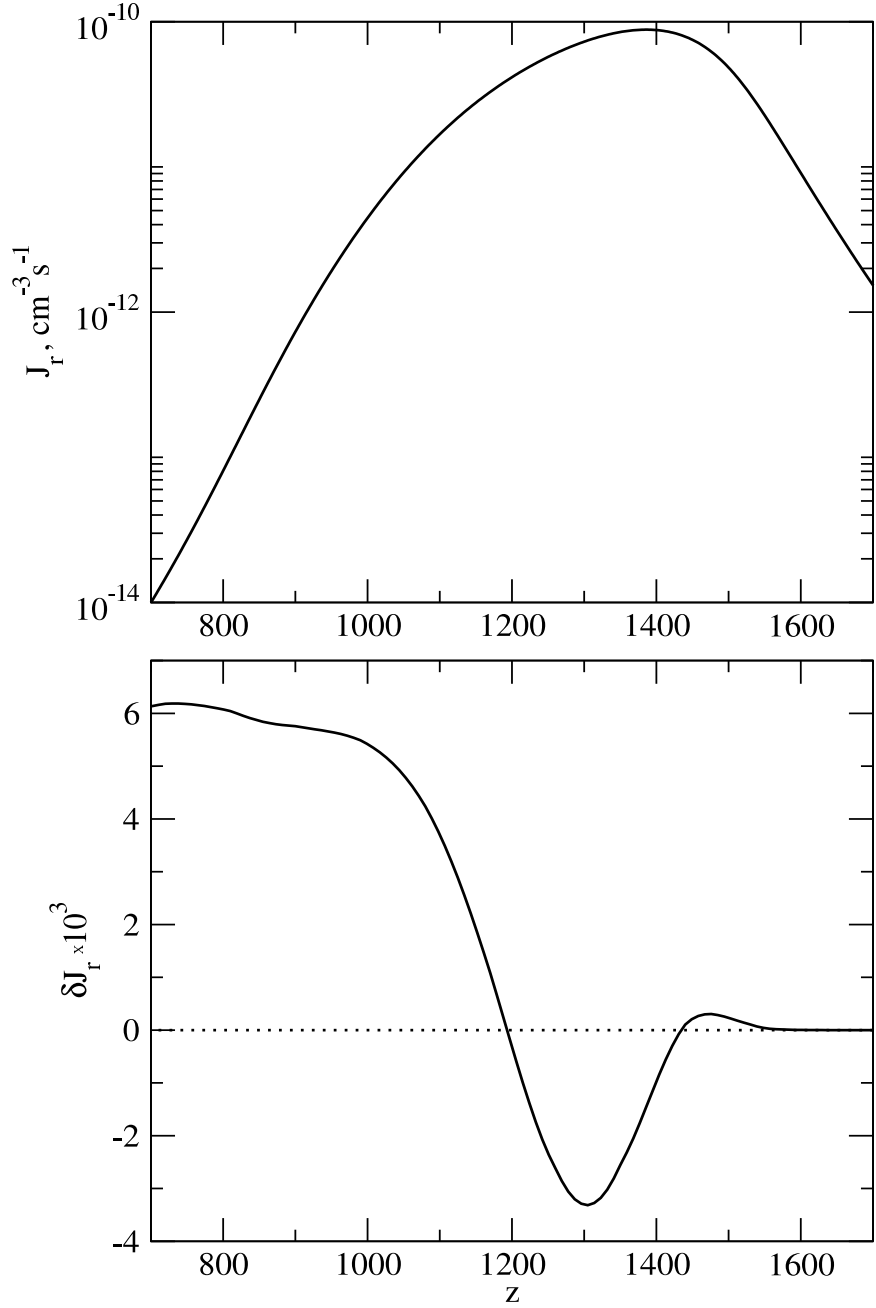}
\caption{Fig. 8. Top: Flow of irretrievably recombining primordial 
hydrogen plasma electrons $J_{r}$. Bottom: The relative change 
in the flow of recombining electrons $\delta J_{r}$ compared to 
the standard recombination model (Peebles 1968; Seager 
et al. 1999) related to the inclusion of induced transitions.}
\label{graph_net_rate}
\end{figure}

\begin{figure}
\centering
\includegraphics[height=20cm, width=16cm]{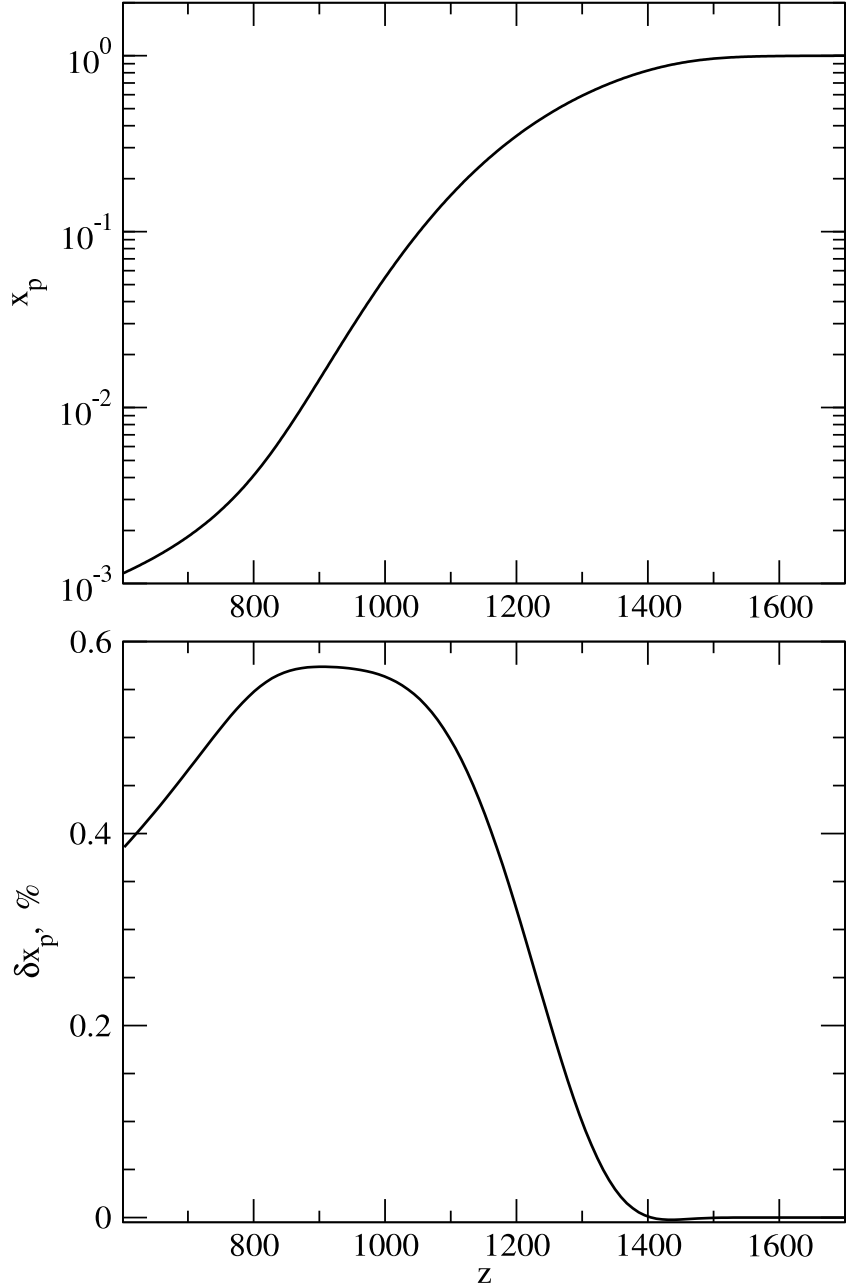}
\caption{Fig. 9. Top: Primordial hydrogen plasma ionization fraction 
$x_p$ vs. redshift $z$. Bottom: The relative change 
$\delta x_p=(x_p-x^{st}_p)/x^{st}_p$ in the ionization 
fraction compared to the standard 
recombination model (Peebles 1968; Seager et al. 
1999) related to the inclusion of induced transitions.}
\label{graph_ion_frac}
\end{figure}
 
\end{document}